\documentclass[reprint,amsmath,amssymb,aps,prb,floatfix]{revtex4-1}

\usepackage{graphicx}
\usepackage{dcolumn}
\usepackage{bm}

\begin{document}

\title{Surface and bulk Landau levels in thin films of Weyl semimetals}

\author{Enrique Benito-Mat\'{\i}as}
\email{enriquebenito2000@gmail.com}
\author{Rafael A. Molina}
\email{rafael.molina@csic.es}
\author{Jos\'e Gonz\'alez}%
 \email{j.gonzalez@csic.es}
\affiliation{%
 Instituto de Estructura de la Materia - CSIC, Serrano 123, E-28006 Madrid, Spain}%

\date{\today}

\begin{abstract}
We show that the thin films of Weyl semimetals have a regime of parameters in which they develop very flat Landau bands under strong magnetic fields. Addressing the case of thin films in a perpendicular magnetic field, we observe that two different types of Landau states may arise depending on whether the line connecting a pair of opposite Weyl nodes is parallel or perpendicular to the direction of the magnetic field. In the latter instance, we show that the flat Landau bands are made of states peaked at the two faces of the thin film. When the line connecting the Weyl nodes is parallel to the magnetic field, we see instead that the states in the Landau bands take the form of stationary waves with significant amplitude across the bulk of the material. In either case, the states in the flat levels are confined along longitudinal sections of the thin film, turning into edge states with distinctive profiles at the lateral boundaries for the two different types of Hall effect.
\end{abstract}

\pacs{Valid PACS appear here}
\maketitle


{\em Introduction.---}
In recent years, we have seen the discovery of different 3D semimetals in which the Fermi surface is made of a number of nodes endowed with topological protection \cite{Wan11,Burkov11,Liu14,Neupane14,Borisenko14,Xu15,Burkov16,Fang16}. The Weyl semimetals (WSs) are unique in its class, as they have Weyl nodes acting like monopole charges of Berry curvature in momentum space. This is at the origin of the so-called chiral anomaly, which leads to the imbalance of the electronic charge in opposite Weyl nodes under suitable electric and magnetic fields.


Recently, the WSs have been investigated in the presence of a strong magnetic field, looking for signatures of the quantum Hall effect \cite{Yang11,Potter14,Lu2015,Cao15,Wang17}. From a theoretical perspective, it has been shown that the 3D nodal-line semimetals may host a rich structure of flat Landau bands, which are topologically protected as long as the particle-hole symmetry is preserved \cite{Molina18}. The discussion of the effect of a strong magnetic field on the 3D WSs is however more delicate, since they afford less symmetry than the 3D nodal-line semimetals. Moreover, the relative orientation of the line connecting opposite Weyl nodes may play an important role in transport properties.

Thus, when the line joining opposite Weyl nodes and the magnetic field are parallel, the Landau bands keep the chiral symmetry of the original model \cite{Lu2015} but, in the case where their directions are perpendicular, chiral symmetry is not preserved and a gap opens due to the mixing of the chiral Landau levels. This destruction of the Weyl nodes has been observed in experiments for the WSs TaP and TaAs at very high magnetic fields fully in the quantum limit\cite{Zhang2017,Ramshaw2018,foot}.  
On the other hand, confirmation of a width dependent quantum Hall transport has only been achieved by a carefully tuned wedge geometry in thin films of the Dirac semimetal Cd$_3$As$_2$ \cite{Zhang2018}, when the angle between the line of nodes and the magnetic field was intermediate between the parallel and perpendicular configurations. 

\begin{figure}[h]
\includegraphics[width=0.70\columnwidth]{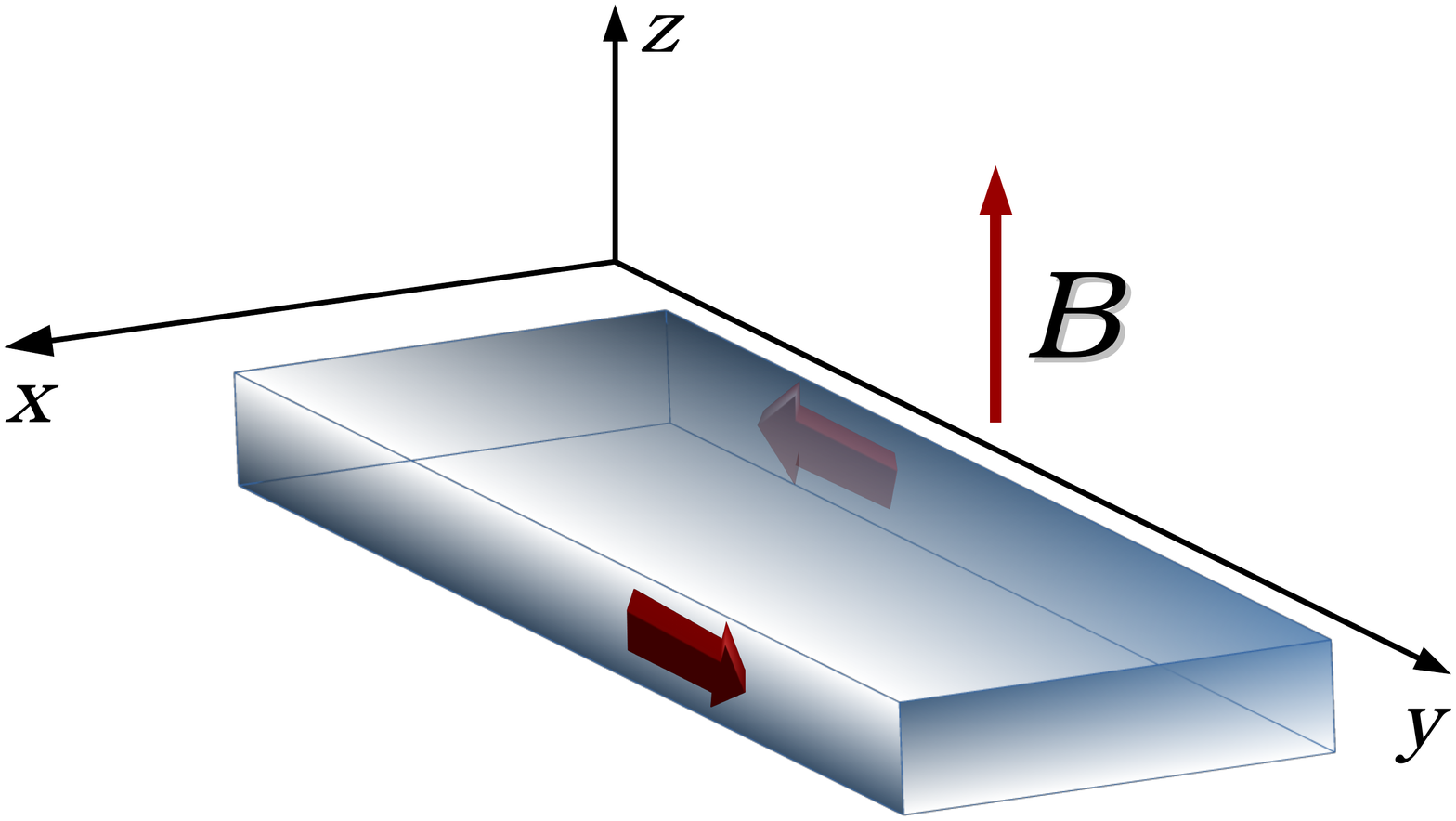}
\caption{\label{wbar} Schematic view of the geometry of the Hall bar considered in the paper, with finite width in the $x$ direction and infinite dimension in the longitudinal $y$ direction.}
\end{figure}

In this paper we identify a regime of parameters of the WSs (actually relevant for the study of the real materials) in which the thin-film geometries develop very flat Landau bands under strong magnetic fields. Addressing the case of thin films with the setup represented in Fig. \ref{wbar}, we observe that two different types of Landau states may arise depending on whether the line connecting a pair of opposite Weyl nodes is parallel or perpendicular to the direction of the magnetic field. 
In the latter instance, where there are Fermi arcs arising from the projection of the line connecting the Weyl nodes onto the surface of the semimetal, we will show that the flat Landau bands are made of states peaked at the two faces of the thin film. These may be considered the counterpart, in a full quantum mechanical picture, of the semiclassical cyclotron orbits connecting opposite Fermi arcs \cite{Potter14,Wang17,Kaladzhyan19}. When the line connecting the Weyl nodes is parallel to the direction of the magnetic field, we will see instead that the states in the Landau bands take the form of stationary waves with significant amplitude across the bulk of the material.

Irrespective of whether the states are attached or not to the faces of the thin film, we will show that they follow a pattern of quantization which is very similar to that of the 2D quantum Hall effect. In the geometry of Fig. \ref{wbar}, that quantization corresponds to the confinement of the states along longitudinal sections, propagating in opposite directions as they approach opposite edges of the bar. Such a localization is indeed behind the existence of very flat Landau bands, which start to disperse only as long as the longitudinal propagation becomes close to the edges of the film, leading to edge states with distinctive profiles at the lateral boundaries for the two different types of Hall effect in the WS.

{\em Landau levels from surface states.---}
We first consider the case in which the line connecting opposite Weyl nodes is parallel to the surface of the thin film. In this setup, that line has a projection onto the two faces of the material, with the consequent formation of surface Fermi arcs. Then, it becomes interesting to find out about the role of the surface states in the quantum Hall regime. In this regard, we are going to see that the low-energy Landau levels in the thin film are not made of Landau states of the 3D material, so that they can be only found by investigating the finite geometry.

We first take the Weyl nodes aligned along the $x$ direction in the geometry of Fig. \ref{wbar}. The magnetic field points in the $z$ direction, and we choose the gauge in which the vector potential is $\mathbf{A} = (0,Bx,0)$. We model then the WS with the Hamiltonian
\begin{eqnarray}
H_{\rm 1} & = & \left[m_0 + m_1 \left( \partial_x^2 - (-i\partial_y+Bx)^2 + \partial_z^2 \right) \right] \sigma_z \nonumber \\ 
 &  & -iv \sigma_x  \partial_z  + v \sigma_y (-i\partial_y + Bx).
\label{hwx}
\end{eqnarray}  
In this case, we can partially diagonalize the Hamiltonian by introducing creation and annihilation operators of the modes of a harmonic oscillator
\begin{eqnarray}
a^{\phantom{\dagger}} = \frac{1}{\sqrt{2}}  \left(\sqrt{B}(x + k_y/B)+\frac{\partial_x}{\sqrt{B}}\right)   \label{ad}   \\
a^{\dagger} = \frac{1}{\sqrt{2}}  \left(\sqrt{B}(x + k_y/B)-\frac{\partial_x}{\sqrt{B}}\right)
\label{ac}
\end{eqnarray}
This brings the Hamiltonian to the form \cite{Sun2019}
\begin{eqnarray}
H_{\rm 1} & = &\left[m_0 + m_1 \left( -2B(a^{\dagger}a^{\phantom{\dagger}}+1/2) + \partial_z^2 \right) \right] \sigma_z  \nonumber \\ 
   &  & - iv \sigma_x  \partial_z + v \sqrt{\frac{B}{2}} \sigma_y  (a^{\phantom{\dagger}}+a^{\dagger}).
\label{hax}
\end{eqnarray}

If we apply the Hamiltonian (\ref{hax}) to the case of the thin film with faces parallel to the $x$-$y$ plane, one can already see that the Landau states cannot take the form of simple stationary waves along the $z$ direction. This becomes clear from inspection of the dependence on $\partial_z $, which makes that the form of the spinor wavefunction vanishing at one of the faces of the film (say at $z = W/2$) cannot match the form of the spinor wavefunction vanishing at the opposite face (say at $z = -W/2$). 

\begin{figure}
\includegraphics[width=0.49\columnwidth]{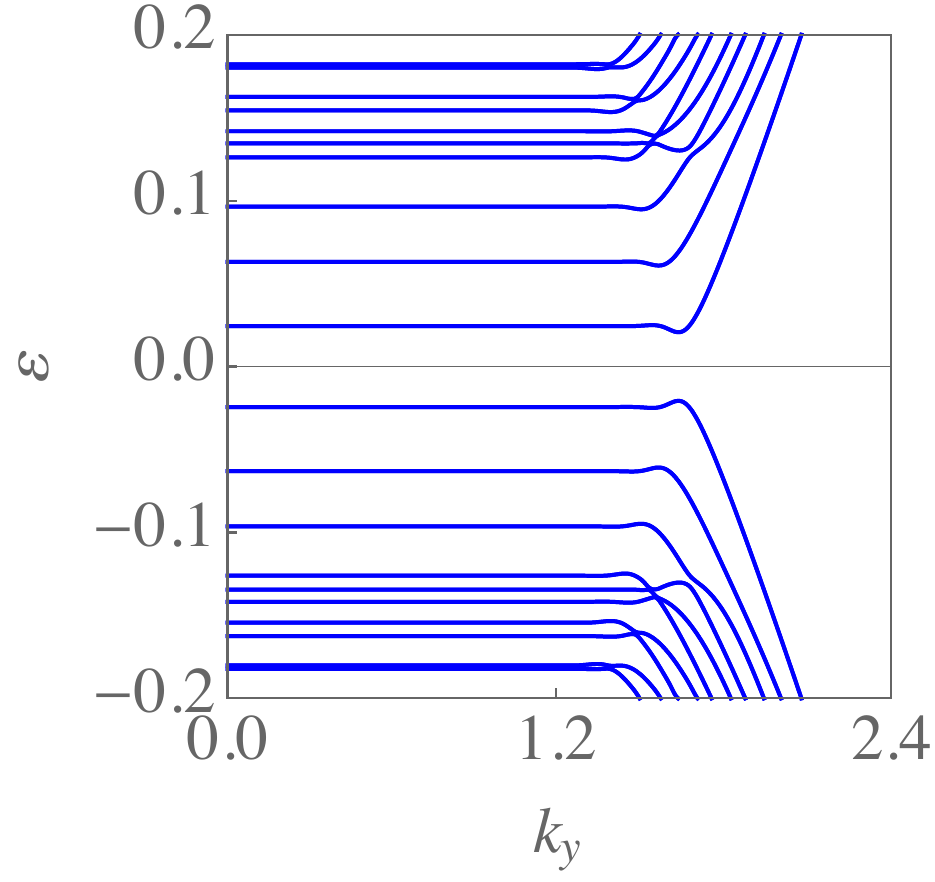}
\includegraphics[width=0.49\columnwidth]{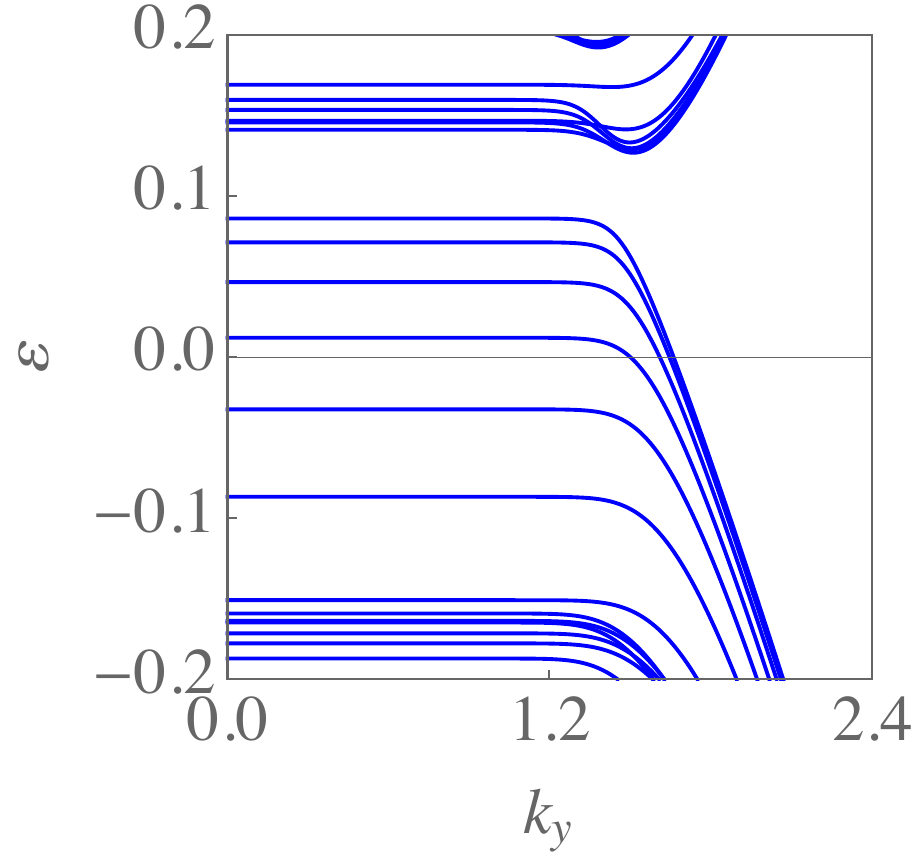}
\caption{\label{levels} Energy bands of a thin film of WS in a perpendicular magnetic field, for two different setups in which the line connecting opposite Weyl nodes is perpendicular (left) and parallel (right) to the direction of the magnetic field. In the two cases the levels have been computed for a bar with transverse dimension $L = 80$ nm and depth $W = 20$ nm under a magnetic field of $B = 30$ T. The parameters used to model the WS are $m_0 = 0.1$ eV, $m_1 = 0.2$ eV nm$^2$, $v = 0.5$ eV nm. The energy is given in eV and the momentum in units of nm$^{-1}$.}
\end{figure}

Yet it can be shown that, for sufficiently strong magnetic field, the eigenstates of the Hamiltonian (\ref{hax}) in the thin film are arranged in flat Landau levels, which develop from the surface states in the Fermi arcs of the WS. In the case of a bar with finite width in the $x$ direction, the bands are flat in a certain range of the momentum $k_y$, but for larger values they become dispersive, as shown to the left in Fig. \ref{levels}, producing a picture that is quite familiar from the 2D quantum Hall effect.

\begin{figure}[h]
\includegraphics[width=0.40\columnwidth]{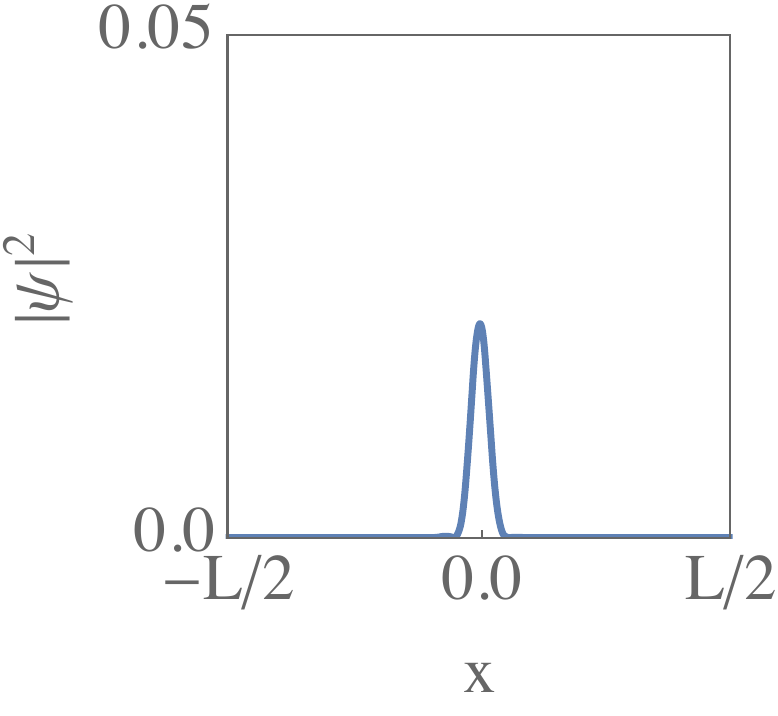}
\includegraphics[width=0.40\columnwidth]{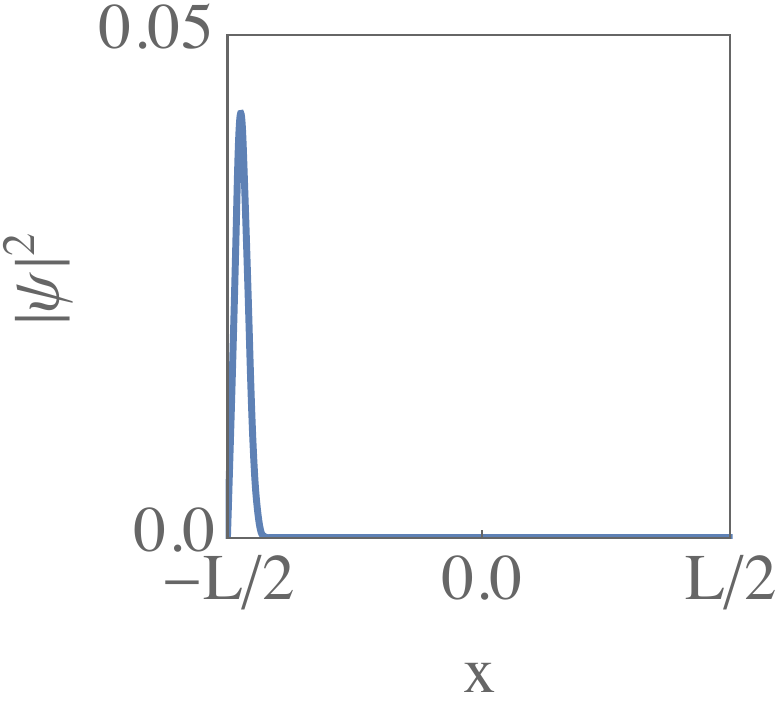}
\caption{\label{statesx_x} Profiles of the probability density along the transverse dimension of a thin film for two different eigenstates in the band with lowest positive energy represented to the left in Fig. \ref{levels}, corresponding to $k_y = 0$ (left) and $k_y = 2.0$ nm$^{-1}$ (right). The profiles are taken at a depth of 3 nm from the upper surface of the thin film.}
\end{figure}

The Landau states follow a clear pattern of localization in the bar of the WS. When $k_y$ is in the flat regime of the Landau level, the corresponding states are localized in longitudinal sections (along the $y$ direction) away from the lateral boundaries of the bar (see Fig. \ref{statesx_x}). As $k_y$ approaches the dispersive regime of the bands, the longitudinal states become progressively closer to the edges of the bar, as shown in Fig. \ref{statesx_x}. 

\begin{figure}[h]
\includegraphics[width=0.40\columnwidth]{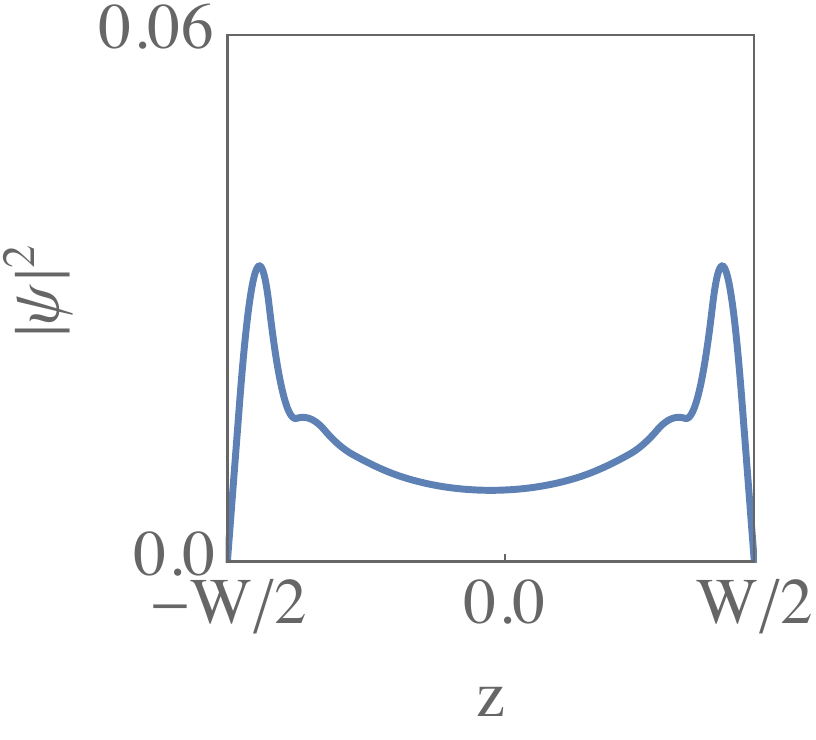}
\includegraphics[width=0.40\columnwidth]{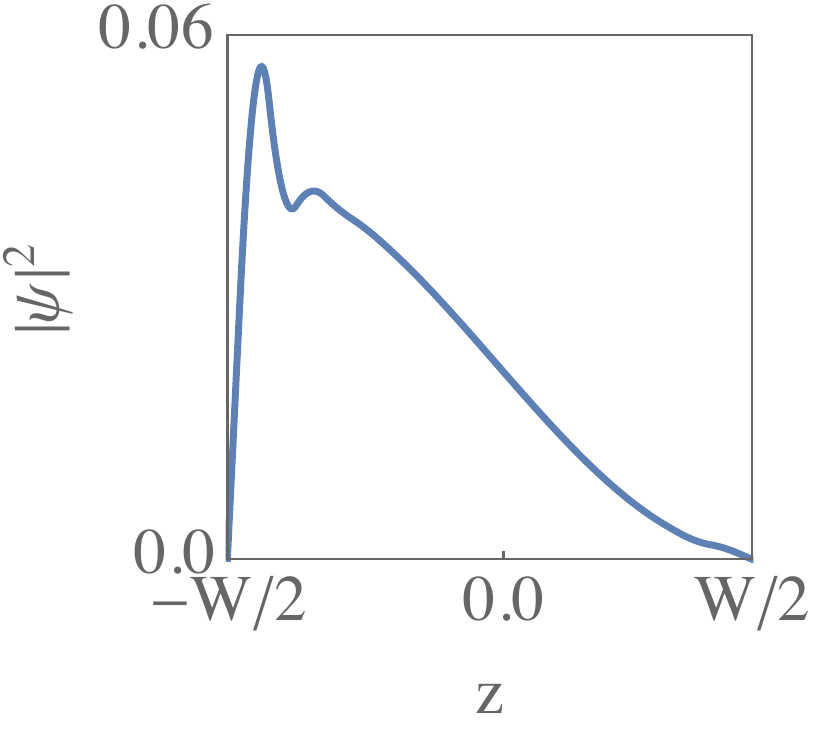}
\caption{\label{statesx_z} Profiles of probability density for the same states considered in Fig. \ref{statesx_x} but taken now along the depth of the thin film, for respective sections at the center of the bar $x = 0$ (left) and at a distance $x$ of 3 nm from the edge (right). The longitudinal dimension of the bar is perpendicular to the line connecting the Weyl nodes as in the above picture.}
\end{figure}

On the other hand, the shape of the probability density along the depth of the film is also quite remarkable. For the states which propagate longitudinally far from the edges of the bar, the probability density is peaked close to each face of the thin film, but with a profile which does not decay completely in the interior of the film (see Fig. \ref{statesx_z}). This distribution changes drastically when the states approach the edges of the bar, since then the probability density accumulates around one of the faces of the film depending on the sign of $k_y$, as shown in Fig. \ref{statesx_z}. 
  
\begin{figure}[h]
\includegraphics[width=0.40\columnwidth]{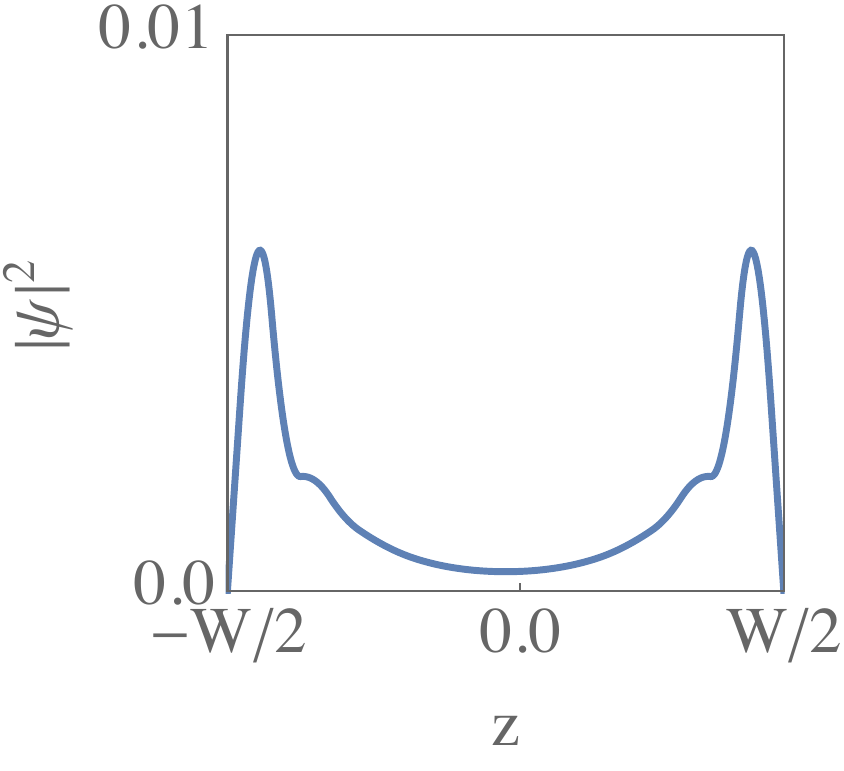}
\includegraphics[width=0.40\columnwidth]{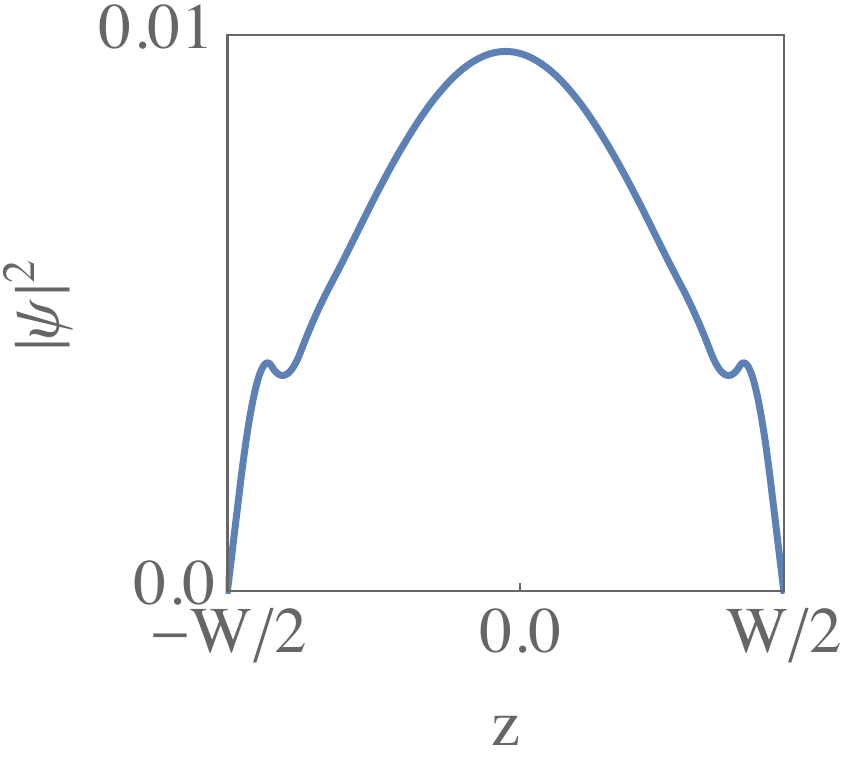}
\caption{\label{statesy_z} Profiles of the probability density along the depth of a bar for two eigenstates of WS with the same geometry and parameters as in Figs. \ref{statesx_x} and \ref{statesx_z}, but with the longitudinal dimension of the bar running parallel to the line connecting the Weyl nodes. The profiles are taken for respective sections at the center of the bar $x = 0$ (left) and at a distance $x$ of 3 nm from the edge (right).}
\end{figure}

The accumulation of the edge states in only one of the faces of the film is indeed characteristic of setups in which the Weyl nodes are not aligned with the longitudinal direction of the bar, which is the case considered in the above discussion. If we change however the orientation of the bar so that its longitudinal dimension becomes parallel to the line connecting the Weyl nodes, then the edge states acquire an even probability density along the depth of the film. In this particular case, the electron density decays drastically inside the film, but it is clearly accumulated in the interior when the electrons run close to the edge of the bar, as can be seen in Fig. \ref{statesy_z}.

{\em Landau levels from bulk states.---}
We discuss now the setup in which the line connecting the Weyl nodes is perpendicular to the faces of the thin film. In this case, the projection of that line onto the surface of the material reduces to a point, so that no contribution to transport can be expected from surface states.

We take the Weyl nodes aligned along the $z$ direction in the setup of Fig. \ref{wbar}, and a magnetic field perpendicular to the thin film with vector potential $\mathbf{A} = (0,Bx,0)$. Then we model the WS with the Hamiltonian
\begin{eqnarray}
H_{2} & = &\left[m_0 + m_1 \left( \partial_x^2 - (-i\partial_y + Bx)^2 + \partial_z^2 \right) \right] \sigma_z \nonumber \\ 
  &  & - iv \sigma_x  \partial_x  + v \sigma_y (-i\partial_y + Bx) .
\label{hwz}
\end{eqnarray}
Making use of the operators (\ref{ad})-(\ref{ac}), we can partially diagonalize the Hamiltonian, which takes the form \cite{Lu2015}
\begin{eqnarray}
H_{2} & = &\left[m_0 + m_1 \left( -2B(a^{\dagger}a^{\phantom{\dagger}}+1/2) + \partial_z^2 \right) \right] \sigma_z  \nonumber \\ 
  & & - i v\sqrt{\frac{B}{2}} \sigma_x(a^{\phantom{\dagger}}- a^{\dagger})+ v\sqrt{\frac{B}{2}}\sigma_y(a^{\phantom{\dagger}}+a^{\dagger}).
\label{haz}
\end{eqnarray}

The main difference with respect to the discussion in the previous section is that now one can find eigenstates of (\ref{haz}) vanishing at both surfaces of the thin film while keeping the same form of the spinor from one face to the other. This means that the wave functions take the simple form of stationary waves inside the film, with well-defined values of the momentum $k_z$ in the $z$ direction. Thus, contrary to the case of the setups considered above, the eigenstates are not attached in the present situation to the faces of the thin film, but they are extended over the interior of the film. This simplifies the resolution of the eigenvalue problem, which can be carried out analytically as reported in Ref. \onlinecite{Lu2015}

\begin{figure}[htb]
\includegraphics[width=0.40\columnwidth]{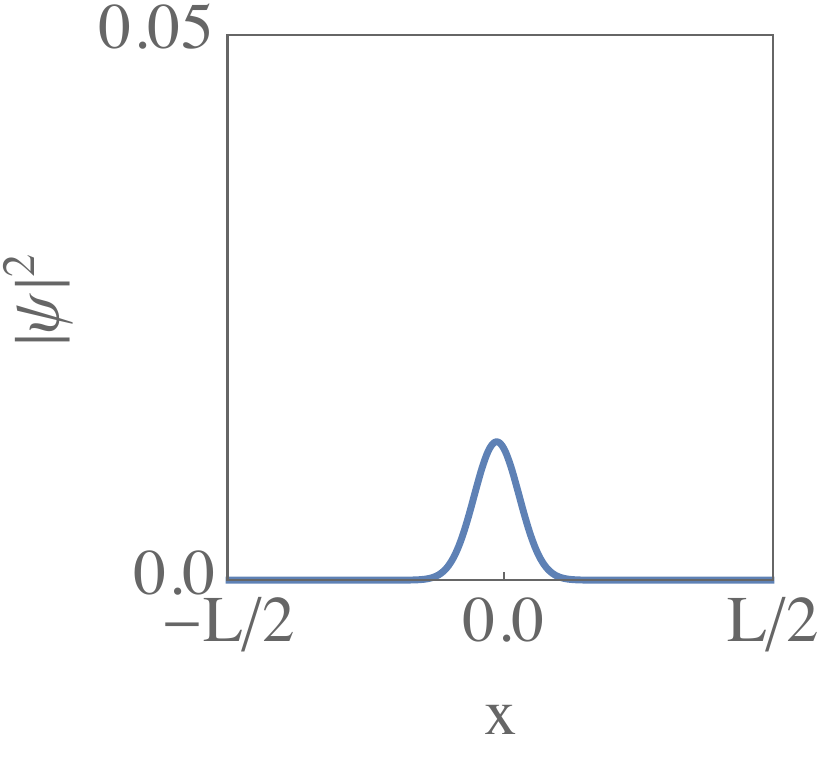}
\includegraphics[width=0.40\columnwidth]{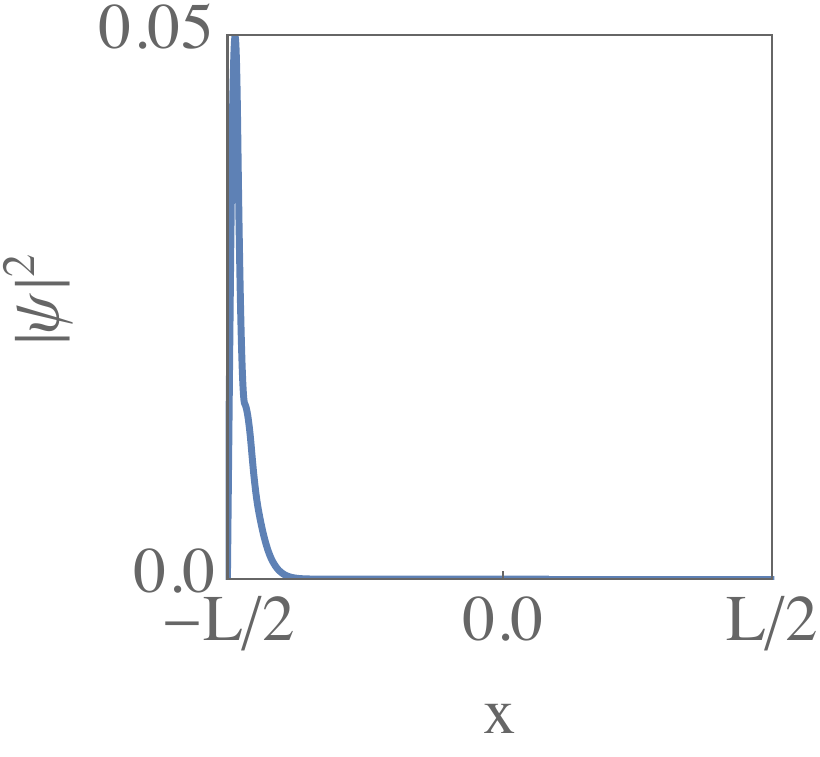}
\caption{\label{statesz_x} Profiles of the probability density for two different eigenstates in the highest hole-like band below the gap in the band structure shown to the right in Fig. \ref{levels}, corresponding to $k_y = 0$ (left) and $k_y = 1.8$ nm$^{-1}$ (right). The profiles are taken along the transverse dimension of the bar, at a depth of 10 nm from the surface.}
\end{figure}

Anyhow, the point we make here is that the bulk character of the eigenstates of (\ref{haz}) does not prevent them from forming flat Landau levels in a sufficiently strong magnetic field. A set of those levels can be seen to the right in Fig. \ref{levels}, which represents the bands of a bar with finite transverse dimension. The flat regime of the bands corresponds to eigenstates that are confined to longitudinal sections away from the edges of the bar. These bands start to get some dispersion as soon as the increase in $k_y$ drives the longitudinal propagation of the states close to the edges of the bar. This change in the localization along the transverse dimension can be seen in Fig. \ref{statesz_x}.

\begin{figure}[htb]
\includegraphics[width=0.40\columnwidth]{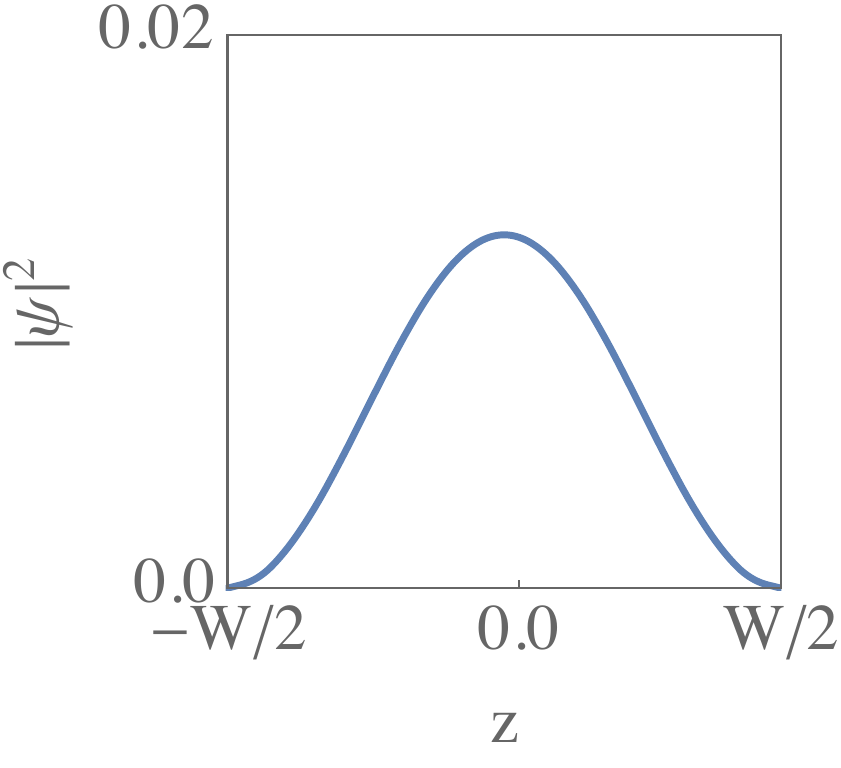}
\includegraphics[width=0.40\columnwidth]{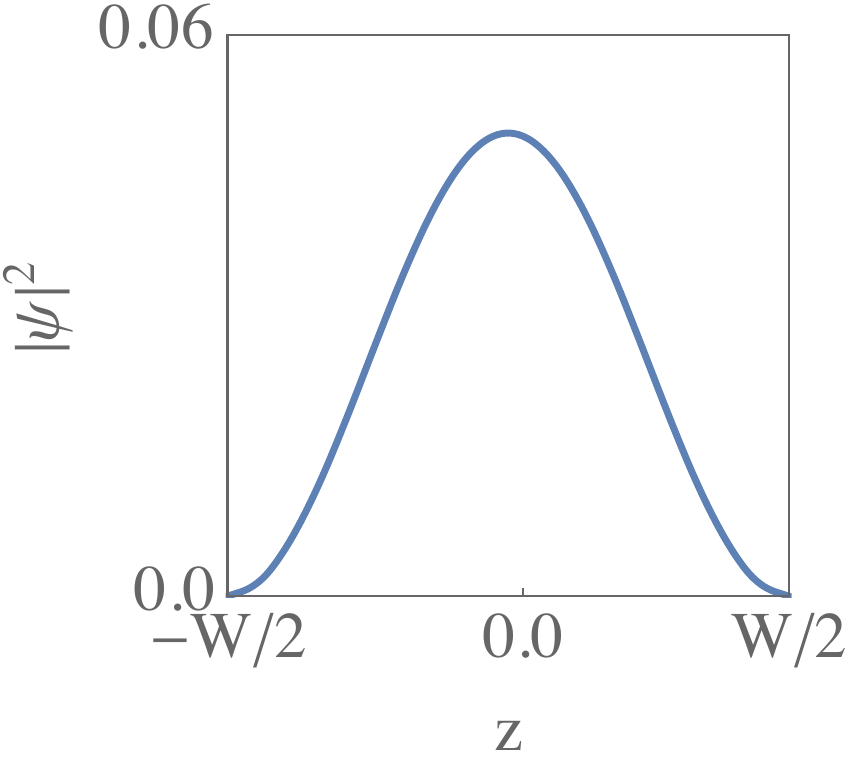}
\caption{\label{statesz_z} Profiles of probability density for the same states considered in Fig. \ref{statesz_x} but taken now along the depth of the thin film, for respective sections at the center of the bar $x = 0$ (left) and at a distance $x$ of 2 nm from the edge (right).}
\end{figure}

In this setup, the probability density of the states in a given Landau band keeps the same shape in the interior of the film, with no much variation as the longitudinal propagation approaches to the edge as shown in Fig. \ref{statesz_z}, which is in accordance with the bulk character of the eigenstates.

{\em Perturbations breaking particle-hole symmetry.---}
The above discussion is based on the hamiltonians (\ref{hwx}) and (\ref{hwz}), which can be taken as a first approximation to model the WSs. This approach captures the low-energy physics of a pair of Weyl nodes, but the description of real materials has to be complemented with the effect of diagonal perturbations (that is, perturbations which are proportional to the identity in pseudospin space).

Real materials displaying WS behavior can be accurately modeled by adding to the hamiltonians (\ref{hwx}) and (\ref{hwz}) new terms which are proportional to the square of the momentum. In order to assess the effect of these perturbations, we pursue then the same analysis carried out before, but adding now to (\ref{hwx}) and (\ref{hwz}) a new term
\begin{equation}
\Delta H  =    a_0 + a_1 \left( \partial_x^2 - (-i\partial_y + Bx)^2 + \partial_z^2 \right) 
\label{hpert}
\end{equation}
The effect of $a_0$ is just to produce a rigid shift of all the energy levels, and it can be disregarded. However, the term introduced by $a_1$ has more significance, in particular because it breaks the particle-hole symmetry of the hamiltonian (\ref{hwx}), while enhancing the deviation from such a symmetry in the hamiltonian (\ref{hwz}).

\begin{figure}[htb]
\includegraphics[width=0.49\columnwidth]{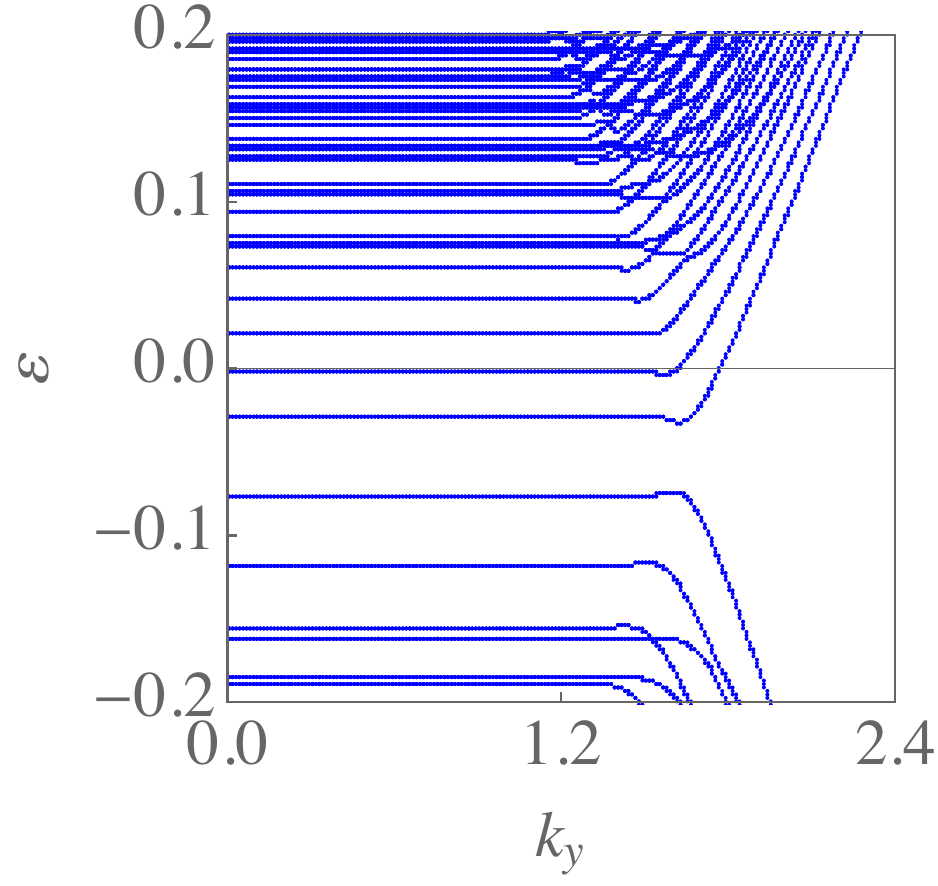}
\includegraphics[width=0.49\columnwidth]{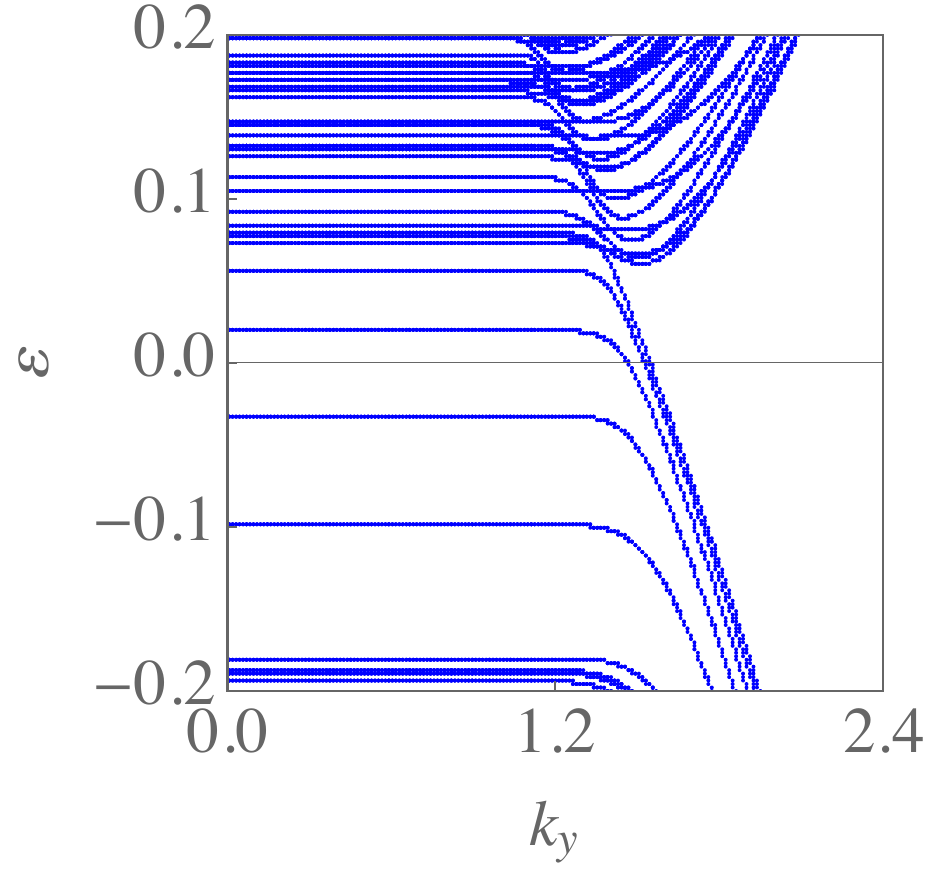}
\caption{\label{pert} Energy bands of a thin film of WS in a perpendicular magnetic field for the same respective setups and parameters considered in Fig. \ref{levels}, with the line connecting opposite Weyl nodes perpendicular (parallel) to the magnetic field for the left (right) plot, but with the addition of diagonal terms in the hamiltonian with $a_0 = 0$ and $a_1 = 0.1$ eV nm$^2$. The energy is given in eV and the momentum in units of nm$^{-1}$.}
\end{figure}

In the case of a thin film with the line connecting the Weyl nodes along the $x$ direction in the setup of Fig. \ref{wbar}, the diagonalization of the hamiltonian $H_1 + \Delta H$ produces the band structure represented to the left in Fig. \ref{pert}. It can be seen that, for sufficiently small values of $a_1$, the original low-energy Landau levels shown to the left in Fig. \ref{levels} can be clearly recognized in the band structure of the perturbed system. However, when $a_1$ goes beyond certain threshold, the Landau levels start to accumulate in a very dense structure in the particle sector (for $a_1 > 0$), already incipient in the plot to the left in Fig. \ref{pert}.

The same effect of condensation of Landau levels can be seen in a thin film with the line connecting the Weyl nodes perpendicular to the surface of the film. In this case, the band structure obtained from the diagonalization of $H_2 + \Delta H$ is represented to the right in Fig. \ref{pert}. The origin of the accumulation of the levels can be more clearly understood in this model, since the momentum $k_z$ is then a good quantum number, accounting for the quantization of the levels with increasing energy. The main role of the perturbation in (\ref{hpert}) is to shift the energy levels downwards (for $a_1 > 0$) by an amount which is proportional to $k_z^2$. This explains that, for a sufficiently large value of $a_1$, the levels start to accumulate in the low-energy regime, producing the picture shown to the right in Fig. \ref{pert}.

{\em Conclusion.---}
We have shown that the thin films of a WS may develop quite flat Landau levels, no matter the orientation of the line connecting the Weyl nodes with respect to the transverse magnetic field. The parameters for which we find in our model very flat bands pertain actually to the experimentally relevant regime realized in materials like TaAs (or Cd$_3$As$_2$ and Na$_3$Bi in the case of the Dirac semimetals). Deviations of the parameters from that regime may however compromise the flatness of the Landau bands, which start to get some curvature as the order of magnitude of $m_0$, $m_1$ or $v$ is changed. 

The Landau levels of the WS are made of states which have 3D support, with a significant extension across the bulk of the thin films. This means that the usual arguments implying the topological protection of the 2D quantum Hall effect cannot be applied in the present context. We stress however that the flatness of the Landau levels may still guarantee the quantization of the Hall conductance, as has been already observed in the parallel configuration \cite{Zhang2018}. In the setup of Fig. \ref{wbar}, the current density $j_y$ along the longitudinal direction is given by the derivative of the energy eigenvalue $\varepsilon $ with respect to the momentum $k_y$. Then, the intensity $I_y$ across the whole transverse section of the bar can be obtained from the expression (reinstating at this point $\hbar $ in the equations)
\begin{equation}
I_y = \frac{e}{\hbar } \int_{\mbox{\rm {\footnotesize filled  states}}}  \frac{dk_y}{2\pi } \frac{\partial \varepsilon }{\partial k_y}
\label{iy}
\end{equation}
The derivative $\partial \varepsilon /\partial k_y $ can be taken as zero except at the dispersive branches of the Landau bands. This means that the integral in (\ref{iy}) is nothing but the difference between the respective chemical potentials $\varepsilon_+, \varepsilon_-$ at the two opposite edges of the bar, times the number $n$ of dispersive bands picked up at the boundaries of the integration. We have therefore that $I_y = n (e/h) (\varepsilon_+ - \varepsilon_-)$, leading to the usual quantization rule of the Hall conductance $G = n(e^2/h)$.

We conclude that the thin films of WSs may provide suitable setups to observe the quantum Hall effect at strong transverse magnetic field. In this picture, the quantization properties rest on the fact that the electronic transport takes place at the edges of the thin film. Our analysis may be useful as a guide for a proper measurement of the edge currents, which may have quite different profiles at the lateral boundaries depending on the orientation of the line of Weyl nodes with respect to the transverse magnetic field.

\acknowledgments

We acknowledge financial support through Spanish grants PGC2018-094180-B-I00 (MCIU/AEI/FEDER, EU), FIS2017-82260-P and FIS2015-63770-P(MINECO/FEDER, EU),  CAM/FEDER  Project  No.S2018/TCS-4342 (QUITEMAD-CM) and CSIC Research Platform PTI-001.


\end{document}